\documentclass[10pt,twocolumn,letterpaper]{article}

\usepackage[T1]{fontenc}
\usepackage{newtxtext,newtxmath}
\usepackage[letterpaper,top=0.66in,bottom=0.72in,left=0.68in,right=0.68in]{geometry}
\usepackage{graphicx}
\usepackage{amsmath}
\usepackage{booktabs}
\usepackage[numbers,sort&compress]{natbib}
\usepackage{stfloats}
\usepackage{tabularx,array,enumitem,placeins}
\usepackage[format=plain,labelsep=period,font=small]{caption}
\usepackage[hyphens]{url}
\usepackage[dvipsnames]{xcolor}
\usepackage{titlesec}
\usepackage[breaklinks,colorlinks,allcolors=MidnightBlue]{hyperref}
\usepackage[capitalize]{cleveref}

\hypersetup{
  pdftitle={Interpreting Speckle Effects Through Neuron Interventions},
  pdfauthor={Shubham K. Singh and Hare Krishna}
}

\renewcommand{\thesection}{\Roman{section}}
\titleformat{\section}
  {\centering\small\bfseries}
  {\thesection.}{0.55em}{\MakeUppercase}
\titlespacing*{\section}{0pt}{1.15em}{0.55em}
\titleformat{\paragraph}[runin]
  {\normalfont\bfseries}{ }{0pt}{}
\titlespacing*{\paragraph}{0pt}{0.65em}{0.45em}

\newcommand{\E}{\mathbb{E}}

\newcommand{\suppref}[1]{Appendix~#1}

\begin{document}

\twocolumn[
\begin{@twocolumnfalse}
\vspace*{0.20in}
\begin{center}
{\fontsize{15.5}{18}\selectfont\bfseries
Interpreting Speckle Effects Through Neuron Interventions\par}
\vspace{0.65em}
{\normalsize
Shubham K. Singh$^{1}$ \; and \; Hare Krishna$^{2}$\par}
\vspace{0.35em}
{\small\itshape
$^{1}$Tufts University; \ $^{2}$University of Texas at Austin\par}
\end{center}

\vspace{0.45em}
\begin{center}
\begin{minipage}{0.88\textwidth}
\small
\noindent
Speckle, the multiplicative noise inherent to synthetic aperture radar (SAR) imaging, reduces the performance of pretrained SAR models. We use interpretability methods to examine its effects inside pretrained vision models. We identify neurons that respond selectively to speckle and test whether correcting their activations improves reconstruction. We study SARMAE, an Earth observation model pretrained on a million-scale SAR dataset. Synthetic scenes with known reflectivity allow us to measure recovery precisely. The model contains 38 speckle-selective neurons that track speckle's value distribution. Restoring these neurons to their clean values does not improve reconstruction. Activation patching across layers shows that the first MLP has the largest restoration and injection effects. Within this layer, a different set of 64 neurons recovers 44.1\% of the correlation loss on held-out scenes. When the speckle-selective neurons are held at their noisy values, 96.5\% of this gain remains, while holding the values of three decoder heads removes 37.4\%. In an ImageNet-pretrained MAE, restoration is most effective in the third MLP block. On manipulated Sentinel-1 and Capella images, the same 64 neurons also improve reconstruction, whereas for land-cover classification, only restoring the entire first MLP yields a clear gain. The neurons that respond most to speckle are therefore not the ones the model uses to recover the scene.

\vspace{0.6em}
\noindent\textbf{Keywords:} Interpretability; Foundation Models; Synthetic Aperture Radar; Speckle; Neuron Interventions

\end{minipage}
\end{center}
\vspace{1.0em}
\end{@twocolumnfalse}
]
\begingroup
\renewcommand{\thefootnote}{*}
\footnotetext{Email: \href{mailto:shubhams12101@gmail.com}{shubhams12101@gmail.com}; \href{mailto:hkrishna.phy@gmail.com}{hkrishna.phy@gmail.com}}
\endgroup

\section{Introduction}
Synthetic aperture radar (SAR) provides observations at night and through cloud cover, supporting applications such as flood mapping and forest monitoring~\cite{moreira2013tutorial}.
One challenge in interpreting these images is speckle, which arises from coherent interference among echoes from many scatterers within each resolution cell and produces intensity fluctuations that can obscure scene structure~\cite{goodman1976speckle}.
Speckle acts multiplicatively on the underlying reflectivity: it changes pixel values while the scene itself stays the same, so the same field can look different in two acquisitions.
Reducing speckle without losing useful spatial detail has motivated decades of research, leading to statistical filters and, more recently, learned despecklers~\cite{lee1980filter,argenti2013speckle,dalsasso2021sar2sar,dalsasso2022merlin}.

Missions such as Sentinel-1 and commercial constellations such as Capella now collect SAR images at a scale that supports foundation models, which are pretrained on large unlabeled images and then adapted to many downstream tasks~\cite{fuller2023croma,guo2024skysense,li2024saratrx,Liu_2026_CVPR}.
Speckle also shapes how these models are built. SAR-JEPA predicts gradient features instead of raw pixels because speckle makes pixel targets unreliable~\cite{li2024predicting}, and SARMAE injects simulated speckle during pretraining so that its features become noise-aware~\cite{Liu_2026_CVPR}.
Whether such choices help is measured by downstream accuracy. Such evaluations quantify how much speckle degrades performance but do not pinpoint the internal mechanisms that drive this loss. Understanding these mechanisms could guide targeted interventions to restore performance.

We ask where to intervene inside a pretrained SAR model to recover scene information degraded by speckle.
Neurons that respond selectively to speckle are natural candidates.
Interpretability methods often characterize neurons by the features that activate them~\cite{bau2017dissection,olah2017feature}, but selectivity need not indicate an effective target for repair.
In image classifiers, for example, a neuron's selectivity is a poor guide to its importance~\cite{morcos2018directions}.
Speckle provides a controlled way to test this distinction: we can generate scenes with known reflectivity, corrupt them according to a known physical model, and measure whether restoring selected neurons improves SARMAE's reconstruction of the underlying scene~\cite{Liu_2026_CVPR}.


We find that the neurons that respond most strongly to speckle mainly follow its skewed value distribution, and restoring them alone does not improve reconstruction.
The recovery comes from elsewhere. Restoring 64 other neurons in the first encoder MLP recovers 44.1\% of the lost correlation on held-out scenes. These neurons carry information about the scene itself: activations from the same scene help, whereas equally large activations from a different scene worsen reconstruction. The recovery also works without the speckle-selective neurons.
When we keep them at their noisy values, 96.5\% of the gain remains, while holding the value vectors of three decoder heads removes 37.4\% of it. A learned corrector can improve reconstruction by predicting these neurons' clean activations from the noisy input, although a simple box filter performs better.
The most effective restoration site also depends on the model: in an ImageNet-pretrained MAE, it shifts to the third MLP.
In SARMAE, the same 64 neurons recover 28.0\% and 35.0\% of the correlation loss on manipulated Sentinel-1 and Capella crops, respectively, when evaluated against smoothed references.
Their benefit does not extend clearly to BigEarthNet-S1 classification, where only restoring the entire first MLP yields a clear gain.
In short, the neurons that respond most to speckle are not the ones this model uses to recover the scene. To decide where to intervene inside a SAR model, we need to test what an intervention restores, not only which neurons respond to the noise.


\begin{figure*}[t]
\centering\includegraphics[width=.85\textwidth]{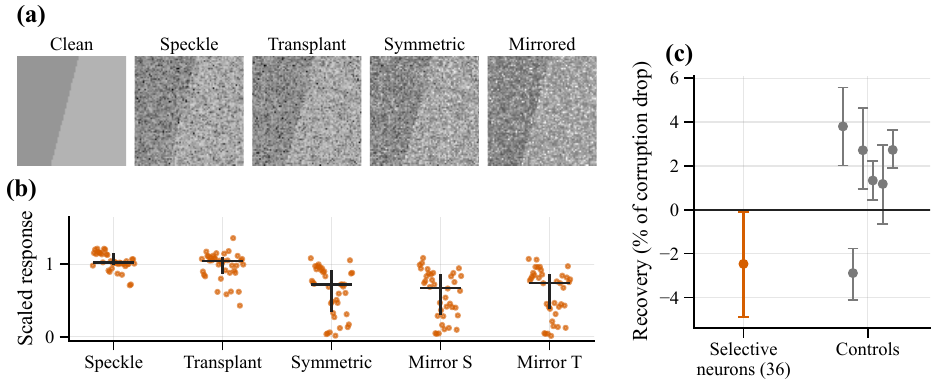}
\caption{\textbf{Speckle-selective neurons track the value distribution, but restoring them does not improve reconstruction.}
(a) Matched stimuli at a grain size of 1.5 pixels separate the effects of value distributions and spatial ranks. The transplant assigns speckle values to another field's ranks, while symmetric and mirrored controls alter the tail shape (Mirror S: mirrored speckle; Mirror T: mirrored transplant).
(b) Scaled responses of the 38 characterized neurons. Each dot shows one neuron's mean response across 24 scenes and two noise draws; black markers indicate medians and interquartile ranges. At coarser grain sizes, responses also depend on spatial statistics.
(c) Restoring only the 36 selective neurons compared with six control sets on the discovery cohort. Recovery is the fraction of the clean-to-noisy correlation drop reversed by restoration; error bars show 95\% scene-bootstrap intervals.}

\label{fig:controls}
\end{figure*}
\section{Related work}
\paragraph{Selectivity and intervention efficacy.}
Network dissection and feature visualization describe what individual neurons respond to~\cite{bau2017dissection,olah2017feature}, and recent work tests these descriptions beyond the most strongly activating examples~\cite{ji2026sieve,oikarinen2026beyond}.
What a neuron responds to and what happens when we change it are different questions.
Morcos et al.~\cite{morcos2018directions} found that class selectivity poorly predicts the effect of ablating a neuron, and Leavitt and Morcos~\cite{leavitt2021selectivity} showed that encouraging class selectivity can hurt generalization.
In language models, Hase et al.~\cite{hase2023localization} found that causal tracing does not reliably identify the best layer for editing a fact.
We ask a related question for a physical corruption, namely whether a successful restoration needs the corruption-selective neurons to change.

\paragraph{Tracing computation.}
Activation patching replaces activations with those from a paired run and measures the effect on the output~\cite{vig2020mediation,heimersheim2024patching}.
Attribution patching estimates these effects with local gradients~\cite{syed2023attribution}, and automated circuit discovery searches for the connections behind a behavior~\cite{conmy2023acdc}.
Which components these methods find depends on how the input is corrupted and how the output is scored~\cite{zhang2024patching}.
In vision, related work has traced influential neuron paths and object-localization mechanisms~\cite{ICLR2025_3e9a55f6,Schaumloffel_2026_CVPR}.
We use attribution patching only to pick candidates and then test them with direct interventions on held-out scenes.

\paragraph{Correcting internal features.}
Other work improves robustness by correcting representations inside the network.
Xie et al.~\cite{Xie_2019_CVPR} add feature-denoising blocks and train them adversarially.
Schneider et al.~\cite{NEURIPS2020_85690f81} replace training-set batch-normalization statistics with statistics from corrupted inputs.
These methods change the model or its statistics.
We keep a pretrained model fixed, ask which of its activations carry recoverable scene information, and then test whether those activations can be predicted from noisy inputs.

\paragraph{Reconstruction and SAR.}
Vision transformers process patch sequences~\cite{dosovitskiy2021vit}, and MAEs learn by reconstructing masked patches~\cite{he2022mae}.
Denoising pretraining improves classification robustness~\cite{wu2023dmae}.
Other studies examine how reconstruction targets affect MAE adversarial robustness~\cite{huang2023frequency} and how reducing sensitivity to corrupted patches makes transformers more robust~\cite{guo2023patchcorruptions}.
SARMAE extends masked reconstruction to SAR imagery~\cite{Liu_2026_CVPR}.
Classical filters and learned despecklers exploit the statistics of speckle~\cite{goodman1976speckle,lee1980filter,argenti2013speckle,dalsasso2021sar2sar}, and MERLIN trains without clean targets by using the real and imaginary parts of complex SAR data~\cite{dalsasso2022merlin}.
Unlike this line of work, we use speckle as a controlled corruption for studying how a fixed pretrained model recovers, and we compare against standard filters.

\begin{figure*}[t]
\centering\includegraphics[width=.90\textwidth]{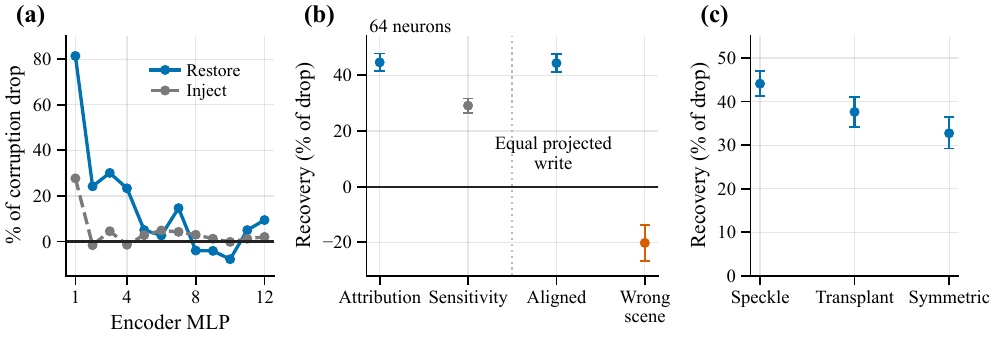}
\caption{\textbf{Restoration depends on the layer, neuron selection, and donor scene.} (a) Mean effects of whole-layer restoration and injection at each encoder MLP on the discovery cohort.
(b) Restoration of 64 MLP1 neurons ranked by attribution or sensitivity on the ranking-confirmation cohort. Clean donors come from the same scene (``Aligned'') or a different scene; interventions are scaled to equal projected write.
(c) Restoration of the 64 neurons across three textures on the restoration-confirmation cohort. Recovery is normalized by each cohort's corruption drop; error bars in (b,c) show 95\% scene-bootstrap intervals.}

\label{fig:restoration}
\end{figure*}

\section{Setup and interventions}\label{sec:methods}
\paragraph{Models and data.}
We use SARMAE's ViT-B/16 encoder and reconstruction decoder.
Each of its 12 encoder MLPs has 3072 post-GELU neurons.
Inputs are $224\times224$ single-channel scenes copied across three channels.
We clip values to $[-40,10]$ dB, map them to $[0,1]$, and normalize with mean and standard deviation 0.5.
As a second model we use the ImageNet MAE-B visualization checkpoint~\cite{he2022mae}.
Our calibration did not favor either 0.5 or ImageNet normalization for this checkpoint, so we report both.
\suppref{A} gives checkpoint identifiers, input and output transforms, and other implementation details.

\begin{figure*}[t]
\centering\includegraphics[width=.9\textwidth]{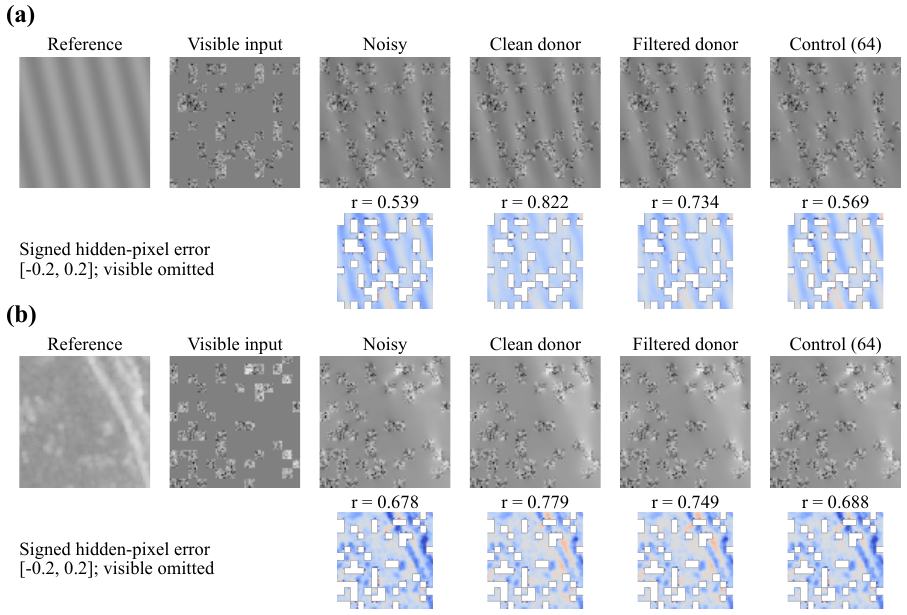}
\caption{\textbf{Reconstruction examples.}
(a) Synthetic stripe scene 34.
(b) Manipulated Capella crop 177.
Each example is closest to its group's median corruption drop (mask 0).
``Clean donor'' restores the 64 selected neurons using activations from the clean reference; for Capella, this reference is a smoothed proxy.
``Filtered donor'' uses activations from a box-filtered copy of the noisy input.
``Control (64)'' applies the same restoration to a 64-neuron control set.
The reported $r$ values are hidden-pixel correlations with the reference.
All intensity images use display limits $[0,1]$; error maps below each reconstruction show signed hidden-pixel errors on $[-0.2,0.2]$.
Visible noisy pixels are identical across reconstructions within each panel.}
\label{fig:examples}
\end{figure*}

For reflectivity $R$, we model speckled intensity as
\begin{equation}
 I=RS,\qquad S\sim\operatorname{Gamma}(L,1/L),
\end{equation}
where $S$ has mean 1 and variance $1/L$.
We control the grain of single-look speckle by filtering complex Gaussian fields before squaring their magnitudes.
In the log domain, speckle has an asymmetric dark tail.
Our statistical controls swap value distributions and spatial ranks between fields, or make the tail symmetric or mirror it (\suppref{B}).
The main experiments use single-look speckle ($L=1$) with 1.5-pixel grain on edges, stripes, and blocks.
Each scene is evaluated with three masks, each showing 49 of 196 patches plus the CLS token.
Our main score is the Pearson correlation $r$ between the reconstructed hidden pixels and the known reference, with contrast gain as a secondary score.
Figure~\ref{fig:examples} shows examples.

\paragraph{Interventions.}
\emph{Restoration} replaces a neuron's noisy post-GELU activation with its value from the clean run.
\emph{Injection} does the reverse.
We call the run that supplies the activations the \emph{donor}.
Unless stated otherwise, the donor uses the same image, mask, and token order, and encoder interventions act on all visible tokens and the CLS token.
A \emph{hold} fixes a downstream signal at its value from the original noisy run while the rest of the model keeps computing.
It tests whether the gain from restoration needs that signal to change.
Clean-donor restoration is an analysis tool, since it needs the clean image.
Filtered donors and learned correction do not.
Filtered donors take activations from a filtered copy of the noisy image, and learned correction predicts clean activations from noisy ones.

Let $r^C,r^S,r^G$ be the clean, noisy, and restored scores, averaged over masks within each scene.
We call $r^C-r^S$ the \emph{corruption drop} and $r^G-r^S$ the \emph{gain} from restoration $G$.
Recovery is the share of the drop that restoration wins back,
\begin{equation}
 R(G)=\frac{\E[r^G-r^S]}{\E[r^C-r^S]}.
 \label{eq:recovery}
\end{equation}
With a hold $H$, the restored score becomes $r^{G,H}$, and the share of the gain that the hold removes is
\begin{equation}
 T(H\mid G)=\frac{\E[r^G-r^{G,H}]}{\E[r^G-r^S]}.
 \label{eq:hold}
\end{equation}

\paragraph{Choosing neurons.}
We call the 36 neurons that stay speckle-selective under masking the \emph{selective neurons} (\cref{sec:selectivity}).
The \emph{restoration neurons} are 64 first-MLP neurons ranked by attribution, the average over discovery scenes of
\begin{equation}
 s_i=\sum_t(a^C_{it}-a^S_{it})\,
 \frac{\partial r^S}{\partial a^S_{it}}.
\end{equation}
For comparison, we also rank neurons in the same layer by \emph{sensitivity}, the absolute difference between their clean and noisy activations.
To fix the set size, we tried 8, 16, 32, 64, 128, and 256 neurons on discovery scenes and kept the smallest set that exceeded half the whole-layer recovery.
This gave 64 neurons, and we did not change the set afterwards.

To compare interventions of different size, we measure an MLP write $\Delta a_t$, the change an intervention makes to a layer's activations at token $t$, by its projected norm $\|W_{\rm out}\Delta a_t\|_2$.
For an exact size match, we scale the larger of two writes down to the smaller norm at every token, and we never scale a write up.
\suppref{D-E} describe both matching schemes.

\paragraph{Cohorts and uncertainty.}
We choose layers, heads, neurons, and set sizes on discovery scenes and test them on new scenes.
Most synthetic cohorts have 72 scenes (24 per family) with three masks each.
Confidence intervals come from resampling scenes within families after averaging over masks.
For ratios, we recompute numerator and denominator together.
Real-image analyses resample sites and acquisitions, and recognition analyses resample geographic tiles.
Each result states its confidence level and any multiple-comparison adjustment.
Table~\ref{tab:cohorts_main} lists the cohorts behind the main results, and \suppref{B} gives the full list.

\begin{table*}[t]
\centering\small
\caption{\textbf{Cohorts used in the main comparisons.}
Counts refer to synthetic scenes (three masks per scene), real-image crops, or recognition patches.
All cohorts were fixed before confirmation.
Appendix~B provides the complete cohort list and documents reuse across experiments.}
\label{tab:cohorts_main}
\begin{tabularx}{\textwidth}{@{}p{5.1cm}lX@{}}
\toprule
Cohort & Count & Role\\\midrule
Discovery / ranking-control discovery & 72 each & Select layers, heads, neuron rankings, and set size.\\
Restoration confirmation & 72 & Recovery of the fixed 64 neurons (44.1\%), ablation, other textures.\\
Ranking confirmation & 72 & Attribution vs.\ sensitivity (44.7\% vs.\ 29.1\%) and donors.\\
Decoder confirmation & 72 & Whole-head, value, and pattern holds.\\
Dependency & 72 & Selective-neuron hold (96.5\% of gain kept) and decoder-value hold.\\
Write / write-budget confirmation & 72 each & Neuron-write controls / decoder-head budget controls.\\
Synthetic fitting / mechanism confirmation & 216 / 72 & Fit / evaluate correctors; GELU hybrids (App.~E).\\
MAE-B discovery / confirmation & 72 each & One checkpoint under two input normalizations.\\
Manipulated Sentinel-1 / Capella & 174 / 70 & Real-image reconstruction against smoothed references.\\
Temporal references & Same 57 & Native-damage test, then added-speckle follow-up.\\
BigEarthNet-S1 train / val. / test & 4,800 / 1,600 / 1,600 & Geographic splits with 32 / 10 / 10 observed tiles.\\
\bottomrule
\end{tabularx}
\end{table*}

\section{What speckle-selective neurons respond to}
\label{sec:selectivity}
We first ask whether these neurons respond to the distribution of speckle values, to their spatial arrangement, or to both.
A response survey found 38 speckle-selective neurons.
To count, a neuron's scaled response to its preferred stimulus had to exceed its response to the most similar non-preferred stimulus (the near-miss control) by at least 0.05, with the 95\% interval of the difference above zero (\suppref{C}).

To separate values from arrangement, we use matched stimuli (Fig.~\ref{fig:controls}A).
The marginal transplant puts speckle-distributed values onto another field's spatial ranks.
The symmetric and mirrored controls keep the decibel mean and spread but change the asymmetry of the distribution (\suppref{B}).
At fine grain, 37 of the 38 neurons prefer both speckle and the transplant over the symmetric and mirrored controls.
Most of the response therefore follows the skewed value distribution, even when the original spatial arrangement is gone.
Deeper neurons prefer finer grain, and at coarser grain the spatial statistics matter more.

With masking, 36 of the 38 neurons still pass the test.
These 36 are the selective neurons used below.
Restoring their clean activations slightly hurts reconstruction, with a recovery of $-2.5\%$ [$-4.9,-0.1$] on the masked discovery cohort (Fig.~\ref{fig:controls}C).

\begin{figure*}[t]
\centering\includegraphics[width=0.72\textwidth]{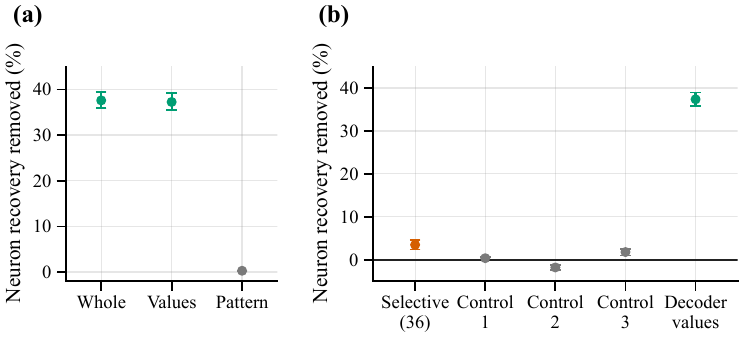}
\caption{\textbf{Restoration gains persist when selective neurons are held fixed but depend partly on decoder values.}
(a) Holds on whole-head outputs, values, or attention patterns in the three selected decoder heads (decoder-confirmation cohort).
(b) Holds on the 36 selective neurons, three control sets matched for neuron count in each layer, or the selected decoder values (dependency cohort).
Both panels report the fraction of the restoration gain removed by each hold (Eq.~\ref{eq:hold}).
Holds are unscaled; error bars show 95\% confidence intervals.
Exact-budget decoder controls are reported in \suppref{E}.}
\label{fig:transmission}
\end{figure*}

\section{What restores scene structure}\label{sec:restoration}
\paragraph{Where restoration works.}
Whole-layer restoration has its largest effect in SARMAE's first encoder MLP, and so does injection (Fig.~\ref{fig:restoration}A).
These scans replace each layer in full, so the size of the intervention differs across layers (\suppref{D}).
Restoring all of MLP1 recovers roughly 80\% on discovery scenes, and our size rule targets half of that.
On the separate restoration-confirmation cohort, the fixed set of 64 neurons recovers 44.1\% [41.2, 47.0].
That is about 2\% of the layer, and the improvement is visible in the hidden patches (Fig.~\ref{fig:examples}).

\paragraph{How the neurons are chosen matters.}
We compare rankings within MLP1 with the number of neurons fixed at 64.
On the ranking-confirmation cohort, the restoration neurons recover 44.7\% [41.5, 47.9] and the sensitivity-ranked neurons 29.1\% [26.5, 31.7], while three random sets recover 0.2-5.1\% (\suppref{D}).
A follow-up in deeper layers compares attribution with selectivity at matched neuron counts: attribution recovers 18.8\% and selectivity $-3.1\%$, a difference of 21.9 [17.7, 26.3] points (\suppref{D}).
Sensitivity measures how much speckle changes a neuron, and selectivity measures how much the neuron prefers speckle over other stimuli.
Attribution instead follows the gradient of the reconstruction score, and testing on new scenes checks that its ranking generalizes.

\paragraph{The gain depends on the scene.}
A large edit to these neurons could help simply because it is large.
We test this by scaling both writes to the same projected norm at every token.
Clean activations from the correct scene then recover 44.4\%, whereas clean activations from another scene of the same family give $-20.2\%$ [$-26.8,-13.8$].
Both interventions target the same neurons and have equal projected write norms at every token.
Only the scene differs, and it decides the sign of the effect.
Controls matched only for write size give smaller gains (\suppref{D}).

\paragraph{The neurons also matter beyond speckle.}
On clean inputs, replacing the 64 neurons' activations with their mean values lowers correlation by 0.042 on the restoration-confirmation cohort.
Six control sets matched by layer and write size lose between $-0.0005$ and 0.0018.
On the same cohort, restoration recovers 37.6\% [34.2, 41.0] for the marginal transplant and 32.7\% [29.2, 36.4] for the symmetric texture, compared with 44.1\% for speckle, and an independent cohort gives similar results (\suppref{D}).

\begin{figure*}[t]
\centering\includegraphics[width=.90\textwidth]{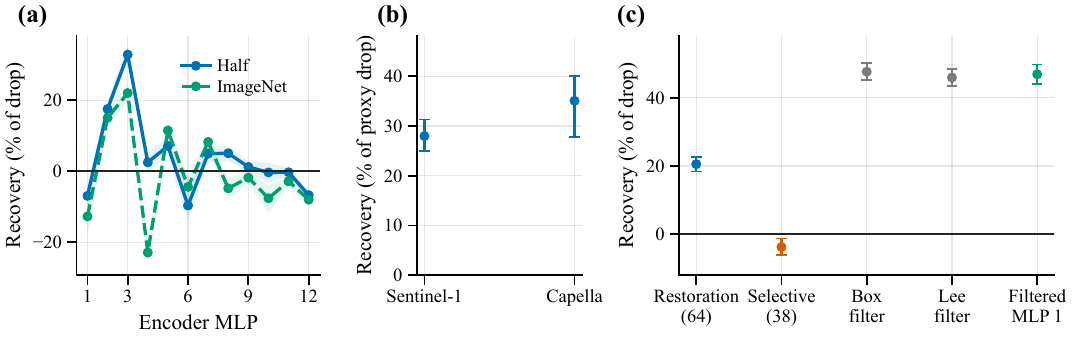}
\caption{\textbf{Scope across checkpoints, SAR images, and correction methods.}
(a) Layerwise restoration in one MAE-B checkpoint under two input normalizations (``Half'': mean and standard deviation both 0.5). Recovery peaks at MLP3.
(b) Restoration of the 64 selected neurons on manipulated Sentinel-1 and Capella crops, evaluated against smoothed references.
(c) Corrections requiring no clean image at inference, evaluated on the mechanism-confirmation cohort: learned corrections to the 64 restoration neurons or all 38 characterized neurons, box and Lee filtering of the full input, and whole-layer insertion of filtered-donor activations at MLP1.
Recovery is normalized by the relevant corruption drop; error bars show 95\% confidence intervals.
Native-image damage and clean-input costs are reported in the text and \suppref{G}.}
\label{fig:scope}
\end{figure*}

\section{What the recovery needs downstream}\label{sec:transmission}
Restoring MLP1 can change downstream encoder and decoder activations.
To see which of these changes matter, we hold selected downstream signals at their noisy values while the rest of the model keeps computing.
The share of the gain that disappears shows how much the recovery depends on that signal changing.

\paragraph{The selective neurons can stay noisy.}
Restoration might still work by changing the selective neurons further downstream.
We test this on a new dependency cohort of 72 scenes with three masks each.
We restore the 64 restoration neurons while holding all 36 selective neurons at their noisy values on every visible token and CLS.
The two sets do not overlap.
A same-layer restoration does not change the three selective neurons in MLP1, so the hold acts on the 33 in MLP2-12.
Per layer, the mean projected norm of the change it undoes ranges from 0.03 to 1.28, or 0.7-28.6\% of the 4.48 norm of the MLP1 restoration (\suppref{E}).

Restoration raises correlation by 0.123.
Holding the selective neurons removes 3.5\% [2.4, 4.6] of this gain and leaves 96.5\% [95.4, 97.6] in place (paired scene bootstrap, Fig.~\ref{fig:transmission}B).
Three control holds with the same number of neurons per layer remove 0.4\%, $-1.8\%$, and 1.8\%.
These controls match neuron counts but not write sizes (\suppref{E}).
A hold keeps the neurons' noisy values, so it blocks their change but keeps their contribution.
Most of the recovery thus happens without the selective neurons moving toward their clean values.

\paragraph{Part of the gain passes through decoder values.}
An attention head outputs $W_O(PV)$, where $P$ is the softmax attention pattern and $V$ holds the value vectors.
A value hold keeps the current pattern but uses the noisy values, a pattern hold does the opposite, and a whole-head hold replaces the head's output with its noisy version.
We chose three heads on discovery data and kept them fixed: block 1 head 14 and block 2 heads 4 and 11 (one-based blocks, zero-based heads; \suppref{E}).

On the decoder-confirmation cohort, whole-head, value, and pattern holds remove 37.6\% [35.9, 39.5], 37.2\% [35.5, 39.2], and 0.3\% [0.2, 0.4] of the gain (Fig.~\ref{fig:transmission}A).
The value hold accounts for nearly all of the whole-head effect.
With write size matched per query, holding these heads removes more of the gain than holding any of the six alternative head sets.
A test of spatial locality did not meet its criterion (\suppref{E}).
On the dependency cohort, the value hold removes 37.4\% [35.8, 39.0] of the gain.

\begin{figure*}[ht]
\centering\includegraphics[width=0.90\textwidth]{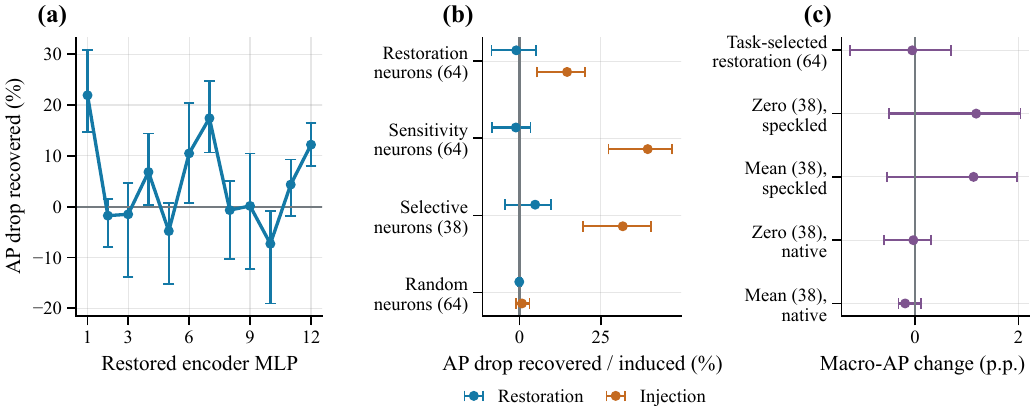}
\caption{\textbf{Whole-layer restoration improves recognition, but sparse restoration shows no clear gain.}
(a) Whole-layer restoration at each encoder MLP using activations from the native input.
(b) Restoration and reverse injection for sparse neuron sets, expressed as a fraction of the macro-AP drop from native to speckled inputs.
(c) Restoration of 64 neurons selected for recognition, and zero or mean ablation of the 38 characterized neurons, reported in raw AP points. Labels indicate the input condition.
Error bars show pointwise 95\% confidence intervals from a paired tile bootstrap.
Multiplicity-adjusted contrasts and matched controls are reported in \suppref{H}.}
\label{fig:recognition}
\end{figure*}
\section{Scope across models, images, and tasks}\label{sec:scope}
\paragraph{A second MAE.}
In the ImageNet-pretrained MAE-B, restoration peaks at MLP3 rather than MLP1 under both input normalizations (Fig.~\ref{fig:scope}A).
Attribution-ranked sets, fixed before confirmation, again beat sensitivity-ranked sets from the same layer.
Set sizes and effect sizes differ between the two normalizations (\suppref{F}).
We ran the selective-neuron and decoder tests only in SARMAE.

\paragraph{Real SAR images.}
We apply speckle to smoothed Sentinel-1 intensity crops and build Capella inputs from a filtered single-look field.
Restoring the 64 neurons recovers 28.0\% [24.9, 31.3] on Sentinel-1 and 35.0\% [27.8, 40.0] on Capella (Fig.~\ref{fig:scope}B).
Both are scored against smoothed references, and the intervals keep geographic and acquisition groups together (\suppref{G}).
We also asked whether native Sentinel-1 GRD images suffer enough speckle damage to test restoration.
Across 57 eligible crops from five sites, temporal references show a correlation drop of 0.0199 [0.0137, 0.0249], below the 0.05 threshold we set in advance.
As a result, we did not run native restoration tests, and our real-image results apply only to controlled manipulations.

\paragraph{Correction without a clean image.}
Because restoration needs the clean image, we also train ridge correctors on synthetic scenes to predict clean activations from noisy ones.
On the mechanism-confirmation cohort, correcting the 64 restoration neurons recovers 20.6\% [18.4, 22.7], while correcting all 38 characterized neurons gives $-3.7\%$ [$-6.0,-1.2$].
The restoration neurons stay useful targets when their clean values must be estimated.
A plain box filter still does better, recovering 47.6\% [45.1, 50.1], and costs less on clean inputs, with a correlation loss of 0.0094 versus 0.0165 for the corrector (Fig.~\ref{fig:scope}C).
\suppref{G} reports whole-MLP correction, filtered donors, Lee filtering, and all clean-input costs.

\paragraph{Recognition.}\label{sec:recognition}
To test a different task, we fit a logistic probe on a fixed SARMAE encoder using refined BigEarthNet-S1 (v2)~\cite{sumbul2021bigearthnetmm,clasen2025reben}.
The probe reaches 47.75 macro-AP on native inputs and 28.22 with added speckle.
With all patches visible, restoring the whole first MLP from the native input raises AP by 4.28 points [2.47, 5.65], recovering 21.9\% [14.7, 30.8] of the drop (Fig.~\ref{fig:recognition}).
Injecting the speckled first-MLP activations into native runs reproduces 100.4\% [92.7, 115.5] of the drop, and whole-layer donors from the wrong scene give $-22.3\%$ [$-44.2,-13.7$].
The tested sparse sets show no recognition gain.
Restoring the 64 restoration neurons changes AP by $-0.19$ points [$-1.37, 0.88$], and the intervals for the sensitivity-ranked and selective sets also include zero.
In fine-tuning, a consistency loss on the 64 neurons gives 44.83 speckled AP versus 44.71 for augmentation alone, and the difference is not statistically resolved.
\suppref{H} reports the task-selected follow-up, ablations, filtering, correction, and training comparisons.

\section{Discussion}\label{sec:discussion}
The neurons that respond most clearly to speckle are not the ones worth correcting.
Restoring them alone slightly hurts, and holding them noisy leaves 96.5\% of the gain from restoring 64 other first-MLP neurons in place.
The useful neurons are ranked by gradient attribution, and the gain they bring depends on scene content.
Our results suggest that activation-level corrections should pick targets by their effect on the task rather than by how strongly they react to the corruption.

The decoder tests locate part of this effect.
Holding the value vectors of three early decoder heads removes about 37\% of the gain, and holding their attention patterns removes 0.3\%.
For these three heads, value holds remove substantially more recovery than pattern holds. 

The practical results are more limited.
A learned corrector for the 64 neurons improves reconstruction without a clean image, but a box filter does better and costs less on clean inputs.
The best layer moves to MLP3 in MAE-B, and targets have to be found for each checkpoint.
For recognition, whole-layer restoration helps, but the sparse reconstruction set does not.
In SARMAE, substantial reconstruction recovery persists when the measured speckle-selective neurons are held at their noisy values. The benefit depends on matching the donor to the scene, even at equal projected write magnitude, and partly on changes in selected decoder values.

\paragraph{Limitations.}
We studied selected neuron sets in two ViT-B checkpoints, and the central dependency tests were run only in SARMAE.
The sets are neither exhaustive nor shown to be minimal.
A noisy hold tests whether a signal must change, but it does not remove the signal.
Layer scans and the different hold types differ in write size, so their effects are not an equal-budget ranking of importance.
The two checkpoints differ in both training data and recipe.
Our real-image results use manipulated inputs and smoothed references.
Native GRD crops show little speckle damage, the results are sensitive to the reference width, and the Capella gain criteria were not met, which limits what we can say about native despeckling (\suppref{G}).

\FloatBarrier
\label{maintextend}
\section*{LLM Usage Statement}
The authors defined the scientific questions, selected the experiments, and determined the methodology.
Claude Code (Anthropic) and Codex (OpenAI) were used as coding assistants for the analyses, figures, and to edit the manuscript for clarity.
The authors directed the work and retain responsibility for the study and all submitted content.
{\small\bibliographystyle{ieeenat_fullname}\bibliography{references}}

\clearpage
\twocolumn[
\begin{center}
{\LARGE\bfseries Supplementary Material}
\end{center}
\vspace{0.6em}
]
\appendix
\setcounter{equation}{0}
\renewcommand{\thetable}{S\arabic{table}}
\renewcommand{\thefigure}{S\arabic{figure}}
\setcounter{table}{0}\setcounter{figure}{0}
\raggedbottom

This supplement gives the model settings, statistical controls, selection rules, and complete scope comparisons. Appendix~D separates the within-layer ranking test from the deeper-layer selectivity comparison. Appendix~E reports the selective-neuron holds, decoder tests, and GELU hybrids. Appendices~F-H give the checkpoint, real-image, correction, and recognition results. 

\section{Models and implementation}\label{app:A}
\paragraph{Checkpoints and preprocessing.}
SARMAE uses the released ViT-B/16 encoder and reconstruction decoder~\cite{Liu_2026_CVPR}: 12 encoder blocks, residual width 768 and 3,072 post-GELU units per MLP; the decoder has eight blocks and 16 heads per block. The optical encoder and alignment module are outside the analyzed route. The converted and original encoder tensors are bit-identical. The reconstruction checkpoint, \path{SARMAE_vitb_checkpoint-last.pth}, records ImageNet-MAE initialization and noise-enabled training with noise ratio 0.5. The second model is the official \path{mae_visualize_vit_base.pth} checkpoint~\cite{he2022mae}, including its original decoder, LayerNorm and exact GELU. 

Scalar displays are repeated over three channels. SARMAE uses mean and standard deviation 0.5, and the fixed $[-40,10]$ dB intensity-to-display window. Predictions are denormalized by branch before channel averaging. MAE-B retains half and ImageNet normalization as two preprocessing branches of one checkpoint. Calibration uses 24 scenes from six families, clean/four-look/training-noise conditions, and three masks. An informative display must beat prediction by the visible-pixel mean. Selecting a normalization requires at least two informative displays, median error ratios above 1.1 and at least 80\% paired wins on each. SARMAE favors half normalization; neither MAE-B branch meets this rule (Table~\ref{tab:calibration_compact}).
\begin{table*}[t]\centering\small
\caption{Normalization calibration. MSE is in $10^{-5}$ display units; the last column gives paired half-normalization wins. The visible-mean criterion rejects the low-gain intensity mapping as uninformative for SARMAE. Both MAE-B normalizations remain in the experiments.}\label{tab:calibration_compact}
\begin{tabular}{@{}llrrr@{}}\toprule
Model & Display / gain & Half MSE & ImageNet MSE & Half wins\\\midrule
SARMAE & Amplitude / 0.1 & 30 & 182 & 99.5\%\\
 & dB / 0.1 & 9 & 150 & 100.0\%\\
 & Intensity / 0.0075 & 1 & 11108 & 100.0\%\\
 & Intensity / 0.1 & 18 & 2076 & 100.0\%\\
MAE-B & Amplitude / 0.1 & 31.30 & 29.78 & 23.6\%\\
 & dB / 0.1 & 24.68 & 20.22 & 25.0\%\\
 & Intensity / 0.0075 & 1.84 & 0.62 & 15.3\%\\
 & Intensity / 0.1 & 10.90 & 9.18 & 18.5\%\\
\bottomrule\end{tabular}\end{table*}

\paragraph{Intervention and token semantics.}
A native MLP computes $a_\ell=g(W_{1\ell}\operatorname{LN}(x_\ell)+b_{1\ell})$ and writes $W_{2\ell}a_\ell+b_{2\ell}$. A neuron is one post-GELU coordinate. Restoration inserts clean activations into a noisy pass; injection reverses the donor and recipient. Donors share image, mask and token order unless specified otherwise. All 49 visible patch tokens and CLS are intervened upon. The decoder restores the 196 patch positions using mask tokens; only hidden pixels are scored. Noisy holds use an independently cached, untouched noisy pass, while subsequent computation remains live.

\paragraph{Metrics and uncertainty.}
For hidden-pixel reference $y$ and reconstruction $\hat y$, the primary correlation and secondary contrast gain are
\begin{align}
r&=\frac{\langle\hat y-\overline{\hat y},y-\bar y\rangle}{\|\hat y-\overline{\hat y}\|\,\|y-\bar y\|},\\
q&=\frac{\langle\hat y-\overline{\hat y},y-\bar y\rangle}{\|y-\bar y\|^2}.
\end{align}
Masks are averaged within each scene before estimation. With clean, noisy, restored and restored-plus-held scores $r^C,r^S,r^G,r^{G,H}$,
\begin{align}
 R(G)&=\frac{\E[r^G-r^S]}{\E[r^C-r^S]},\\
 T(H\mid G)&=\frac{\E[r^G-r^{G,H}]}{\E[r^G-r^S]}.
\end{align}
These are ratios of mean differences: reconstruction recovery divides by the corruption drop; a hold divides by induced recovery. Both means are recomputed together in each bootstrap draw. Neither ratio decomposes total error additively.

Synthetic intervals resample paired scenes within family after mask averaging. Ordinary arm intervals are 95\%; adjusted comparisons are identified locally. Ratios require their stated denominator criterion. Sentinel analyses preserve site or acquisition-site-crop grouping, Capella region-acquisition-crop grouping, temporal-reference analyses site-crop grouping, and recognition MGRS-tile grouping. The 216 masks in a standard cohort are not 216 independent scenes.

\section{Inputs, statistical controls and cohorts}\label{app:B}
Reflectivity $R$ and speckle obey $I=RS$, $S\sim\operatorname{Gamma}(L,1/L)$, with mean one and variance $1/L$~\cite{goodman1976speckle}. Correlated speckle filters independent real/imaginary Gaussian fields before squaring; averaging independent looks varies $L$. Single-look log speckle has a dark tail and mean shift near $-2.51$ dB. The marginal transplant puts speckle-distributed values on another field's spatial ranks. Symmetric and mirrored controls change distributional shape while sharing the specified decibel mean and spread. Higher-order spatial statistics are not all matched. Within-patch shuffling preserves histograms while removing grain organization.

The response library includes smooth/structured scenes, bright points, pure darkening, independent speckle at 1/3/16 looks, correlated single-look grain at 0.75/1.5/3 pixels, four-look correlated speckle, matched lognormal granules, Gaussian display noise and patch-shuffled speckle. White display noise has standard deviation 0.15; released training noise uses lognormal $\sigma=0.5$. Reconstruction uses known-reflectivity edges, stripes and blocks. Each standard cohort has 24 scenes per family and three masks revealing 49/196 patches. Each mask leaves at least two visible patches above pattern mean 0.5 and two below $-0.5$, plus at least 256 hidden pixels of each sign. All arms share masks.

\begin{table*}[t]\centering\small
\caption{Independent cohorts and stated reuse. Synthetic counts are scenes, real-image counts are crops, and recognition counts are patches. Standard synthetic cohorts use three masks per scene.}\label{tab:cohorts_supp}
\begin{tabularx}{\textwidth}{@{}p{5.2cm}lX@{}}\toprule
Cohort & Count & Role\\\midrule
Discovery / ranking-control discovery & 72 each & Layer/head scans, rankings and intervention sizes.\\
Restoration / write / ranking confirmation & 72 each & Neuron effects, ablation, controls and donors; grain variants reuse ranking geometry.\\
Geometry-shift / additional confirmation & 72 each & Curved scenes / location, basis, encoding and filtered donors.\\
Decoder / write-budget confirmation & 72 each & Component holds / exact-budget controls.\\
Dependency cohort (new draws) & 72 & Selective-neuron holds and decoder comparator.\\
Locality-message cohort & 72 & Source/query and matching tests.\\
Synthetic fitting / mechanism confirmation & 216 / 72 & Correction fitting / GELU and correction evaluation.\\
MAE-B discovery / confirmation & 72 each & One checkpoint, two normalization branches.\\
Manipulated Sentinel-1 / Capella & 174 / 70 & 16 sites and 15 acquisitions / 6 regions and 8 acquisitions.\\
Temporal references, native / added speckle & Same 57 & 5 of 6 planned sites; second experiment reuses references and masks.\\
BigEarthNet-S1 training / validation / test & 4,800 / 1,600 / 1,600 & 32 / 10 / 10 observed tiles.\\
\bottomrule\end{tabularx}\end{table*}

Sets are fixed on discovery scenes before any confirmation outcome is computed. New draws test the fixed interventions; comparisons that reuse scene geometry or previously inspected examples are marked as such. Designs were written down locally before each measurement but were not externally preregistered, and follow-ups motivated by earlier outcomes are labeled as follow-ups. 


\section{Neuron selection and response characterization}\label{app:C}
The response survey measures 36,864 MLP coordinates on 4,200 images. Taking three neurons per layer under each of four rankings enriches strong responders for manual interpretation. Of these, 38 are characterized as speckle-selective; 36 retain selectivity under masking. Validation requires a driver-minus-near-miss scaled response of at least 0.05 and a positive lower 95\% bound. This procedure does not estimate population monosemanticity.

The 38 response-selected neurons, 36 masked-retained neurons, 64 sensitivity-ranked neurons and 64 restoration neurons are distinct sets. The latter are selected by reconstruction attribution (Section~\ref{app:D}) and have no overlap with the 36 selective neurons. Their joint intervention therefore has no coordinate conflict.

At 1.5-pixel grain, median scaled responses are 0.49 (clean), 1.02 (speckle), 1.05 (transplant), 0.72 (symmetric), 0.68 (mirrored speckle) and 0.75 (mirrored transplant); 37/38 neurons pass the joint preference test. Spatial statistics matter more at coarse grain. Across all nine recorded effect/margin thresholds, depth versus preferred-grain correlations range from $-0.20$ to $-0.54$, with coarser preferences more frequent than fine ones. On real images, 25 of 27 response predictions hold and two broad-noise comparisons tie. In the earlier description test, one primary prediction tied. In a stricter post hoc descriptive check, every driver-family mean exceeds every near-miss-family mean for 18 of the 38 speckle-selective neurons.

Restoring the 36 masked-retained neurons alone gives $-2.5\%$ [$-4.9,-0.1$] recovery on discovery scenes. This estimates a standalone intervention; the separate hold experiment tests dependence of the induced recovery on their changes.

\section{Restoration, donors and supporting controls}\label{app:D}
Attribution ranks the discovery mean of $(a^C-a^S)\partial r^S/\partial a^S$, summed over visible tokens and CLS. Sensitivity uses mean absolute paired activation change. The attribution-sensitivity comparison keeps the layer fixed at MLP1 and compares equal neuron counts; it tests target selection within one location. Actual restoration tests sizes 8, 16, 32, 64, 128 and 256; the smallest tested set exceeding half of whole-MLP recovery contains 64 neurons. With roughly 80\% whole-layer recovery on discovery scenes, this rule selects for recovery on the scale of 40\%. The 44.1\% recovery reported on restoration confirmation is a separate measurement of the fixed set on new scenes. Whole-component scans are not magnitude matched across layers. Table~\ref{tab:ranking_compact} gives the paired ranking and donor comparisons, without substituting estimates from another cohort.
\begin{table*}[t]\centering\small
\caption{Ranking-confirmation cohort: correlation-drop recovery (\%) with 95\% intervals. All rankings are selected on separate discovery scenes. Balanced donors write into the same 64 restoration neurons and are attenuated to equal projected norm at every token; the wrong donor comes from a different scene in the same family.}\label{tab:ranking_compact}
\begin{tabular}{@{}lrr@{}}\toprule
Ranking & 32 neurons & 64 neurons\\\midrule
Attribution & 32.4 [29.5,35.3] & 44.7 [41.5,47.9]\\
Sensitivity & 21.5 [19.3,23.9] & 29.1 [26.5,31.7]\\
Activation & 5.9 [4.7,7.1] & 13.4 [11.6,15.2]\\
Projected-write proxy & 2.1 [1.1,3.1] & 5.1 [3.6,6.6]\\
Random set 1 & - & 0.2 [0.0,0.4]\\
Random set 2 & - & 4.6 [4.0,5.1]\\
Random set 3 & - & 5.1 [4.4,5.8]\\\midrule
Donor comparison & \multicolumn{2}{c}{64 restoration neurons}\\\midrule
Discovery-clean mean & \multicolumn{2}{c}{-34.4 [-38.3,-30.6]}\\
Wrong scene, unattenuated & \multicolumn{2}{c}{-32.3 [-42.3,-23.1]}\\
Correct scene, balanced & \multicolumn{2}{c}{44.4 [41.2,47.6]}\\
Wrong scene, balanced & \multicolumn{2}{c}{-20.2 [-26.8,-13.8]}\\
\bottomrule\end{tabular}\end{table*}

The applied MLP write is $w_t=W_2\Delta a_t$. Summed per-coordinate norms are only a selection proxy because contributions can cancel. Exact-budget pairs instead attenuate both writes to the smaller $\|w_t\|$ at every token, without amplification. Correct and wrong-scene clean donors from the same family thus differ in scene information at equal applied magnitude.

On the restoration-confirmation cohort, mean ablation of the 64 restoration neurons costs 0.0421 correlation on clean input; six layer/write-proxy-matched controls range from $-0.0005$ to 0.0018. The separate write-confirmation cohort has cost approximately 0.039. Cross-texture recovery is 44.1 [41.2,47.0]\% (speckle), 37.6 [34.2,41.0]\% (transplant) and 32.7 [29.2,36.4]\% (symmetric) on restoration confirmation, versus 43.8 [40.8,46.8]\%, 38.7 [35.7,41.7]\% and 31.3 [27.8,35.0]\% on decoder confirmation.

\paragraph{Selection outside MLP1.}
The initial location-by-criterion comparison includes MLP1 in its distributed set. A follow-up excludes MLP1 and compares attribution with selectivity at matched neuron counts in deeper layers. Attribution recovers 18.8\%, compared with $-3.1\%$ for selectivity, a difference of 21.9 [17.7,26.3] recovery percentage points. This comparison addresses selection outside MLP1. It is separate from the within-MLP1 attribution-sensitivity comparison: selectivity measures response preference, while sensitivity measures the size of the clean-noisy activation change.

\paragraph{Additional restoration controls.}
The fixed restoration set recovers 43.4\%, 28.8\% and 43.5\% under fine grain, coarse grain and shifted geometry. Native coordinates outperform fitted PCA/random directions at equal projected norm, while unattenuated comparisons differ. Encoding probes classify 11/64 units as local-mean related, 14 as edge/orientation related and 39 as neither; the stronger structural-carrier criterion is not met. Signed intensity associations remain exploratory. 

Examples are selected from baseline corruption drops alone, averaged over masks: the scene/crop nearest the specified cohort median, tie by ID, with mask zero. Visible pixels retain observations, and all reported scores use hidden pixels. Reference and error scales are fixed; proxy references are identified.

\section{Dependency and decoder transmission}\label{app:E}
\paragraph{Dependency experiment.}
A new cohort contains 72 scenes (24 per family), three masks each, and independent scene/mask/corruption seeds. The 64 restoration neurons and 36 selective neurons are the same fixed sets used above. The latter have layer counts 3,3,4,3,4,4,3,3,3,3,1,2 across MLP1-12 and are disjoint from the restoration set.

Untouched clean and noisy passes cache paired post-GELU activations. A single hook per layer first inserts clean values for the 64 restoration neurons at MLP1, then holds selective coordinates to baseline-noisy values, including CLS and all visible tokens. Other computation remains live. Three fixed controls exclude both target sets and match selective-neuron counts at every layer. They are unattenuated and \emph{not write-norm matched}. The three selective neurons in MLP1 are not changed by a same-layer restoration, so their hold applies no write (recorded norm 0); the hold is active for the 33 selective neurons in MLP2-12. There, mean projected hold-write norms per layer range from 0.03 (MLP2) to 1.28 (MLP7), against 4.48 for the MLP1 restoration write. Dividing the layerwise mean hold norms by this reference gives 0.7-28.6\%; these ratios compare different layers and are descriptive. These recorded writes show that the holds reverse nonzero changes in the downstream selective set. They measure the size of the projected intervention, not the fraction of the clean-noisy activation gap closed by restoration. The positive comparator holds noisy values in the previously selected decoder trio while leaving patterns live.

The primary effect is $\E[r^{64}-r^{64,H}]$ after mask averaging; its normalized form divides by $\E[r^{64}-r^S]$. Ten thousand paired family-stratified scene resamples produce 95\% percentile intervals, recomputing both means per draw. The clean/noisy/restored scores are 0.9513 [0.9425,0.9589], 0.6780 [0.6521,0.7021] and 0.8014 [0.7778,0.8222]. The corruption drop is 0.2733 [0.2539,0.2937]; restoration adds 0.1234 [0.1142,0.1325], or 45.13 [42.13,47.83]\% of that drop.

\begin{table*}[t]\centering\small
\caption{Dependency cohort: unattenuated holds during 64-neuron restoration. Fractions divide by induced recovery, not the corruption drop. Controls match layer counts only; their intervals are descriptive. Negative loss denotes improvement.}\label{tab:selective-hold}
\begin{tabular}{@{}lrr@{}}\toprule
Held component & Correlation loss [95\% CI] & Recovery removed (\%) [95\% CI]\\\midrule
36 selective neurons & 0.0043 [0.0030,0.0057] & 3.48 [2.42,4.60]\\
Layer-count control 1 & 0.0005 [0.0000,0.0009] & 0.37 [0.00,0.72]\\
Layer-count control 2 & -0.0022 [-0.0030,-0.0015] & -1.79 [-2.40,-1.21]\\
Layer-count control 3 & 0.0022 [0.0011,0.0033] & 1.80 [0.97,2.56]\\
Decoder values & 0.0461 [0.0428,0.0495] & 37.37 [35.81,38.99]\\
\bottomrule\end{tabular}\end{table*}

Holding the selective set preserves 96.52 [95.40,97.58]\% of the induced recovery and removes a small positive part of the gain (Table~\ref{tab:selective-hold}). The result establishes that substantial recovery survives with these measured responses fixed at their noisy values. The hold retains their noisy-state contribution and tests dependence on their restoration-induced changes. It does not test removal of the neurons or cover every corruption-responsive representation.

Noisy self-holds reproduce baseline outputs within $6.3\times10^{-7}$. Restoration and combined-hook ordering agree bitwise with independent implementations; the decoder comparator agrees within $5.4\times10^{-7}$. Actual projected writes are recorded per layer/token: $\|W_{\mathrm{out},S}(a^{\mathrm{donor}}_S-a^{\mathrm{current}}_S)\|_2$, with decoder values measured per query. The primary selective hold is not attenuated. Per-example records, norm summaries and checks accompany the experiment.

\paragraph{Decoder components and magnitude controls.}
For a head, output is $W_O(AV)$, $A=\operatorname{softmax}(QK^\top/\sqrt d)$. A value hold uses the current pattern with baseline-noisy values; a pattern hold uses the noisy pattern with current values; a whole-head hold inserts its noisy output. Later blocks respond to earlier holds. The discovery-selected trio is L1H14, L2H4 and L2H11 (one-based blocks, zero-based heads). On decoder confirmation, whole-head/value/pattern holds remove 37.6 [35.9,39.5]\%, 37.2 [35.5,39.2]\% and 0.3 [0.2,0.4]\% of induced recovery. Values similarly dominate for transplant and symmetric textures (Table~\ref{tab:decoder_components}).
\begin{table*}[t]\centering\small
\caption{\textbf{Decoder whole-head, value and pattern holds by texture.} Entries are percentages of neuron recovery removed; brackets give 95\% confidence intervals.}\label{tab:decoder_components}
\begin{tabular}{@{}p{3cm}p{4.3cm}p{4.3cm}p{4.3cm}@{}}
\toprule Texture & Whole heads & Values only & Patterns only\\\midrule

Speckle & 37.6 [35.9, 39.5] & 37.2 [35.5, 39.2] & 0.3 [0.2, 0.4] \\
Transplant & 38.4 [36.7, 40.1] & 38.1 [36.4, 39.8] & 0.3 [0.2, 0.4] \\
Symmetric & 43.5 [41.4, 45.5] & 43.0 [40.8, 45.0] & 0.4 [0.3, 0.5] \\
\bottomrule\end{tabular}
\end{table*}
\begin{table*}[t]\centering\small
\caption{\textbf{Equal-budget decoder-head comparisons.} Each control lists one head in decoder block one and two heads in decoder block two. Scale values are listed for blocks one and two. The selected trio is block one head 14 and block two heads 4 and 11. Intervals are Bonferroni-adjusted 99.1667\%. }\label{tab:head_budgets}
\begin{tabular}{@{}p{5.6cm}rrrr@{}}
\toprule Control heads & Selected loss & Control loss & Difference [CI] & Selected scale\\\midrule

Block one head 3; block two heads 0, 10 & $0.0358$ & $0.0018$ & $0.0340\ [0.0307, 0.0374]$ & 0.922/0.977 \\
Block one head 2; block two heads 0, 10 & $0.0347$ & $0.0020$ & $0.0327\ [0.0295, 0.0359]$ & 0.893/0.976 \\
Block one head 8; block two heads 0, 10 & $0.0364$ & $0.0018$ & $0.0347\ [0.0312, 0.0381]$ & 0.930/0.977 \\
Block one head 0; block two heads 0, 10 & $0.0327$ & $0.0020$ & $0.0307\ [0.0275, 0.0338]$ & 0.852/0.975 \\
Block one head 4; block two heads 0, 10 & $0.0323$ & $0.0031$ & $0.0293\ [0.0263, 0.0322]$ & 0.804/0.975 \\
Block one head 2; block two heads 0, 13 & $0.0311$ & $0.0043$ & $0.0268\ [0.0242, 0.0295]$ & 0.893/0.929 \\
\bottomrule\end{tabular}
\end{table*}

Exact projected budgets are matched per query and separately by decoder block. Each of six control pairs therefore has its own attenuated selected intervention (Table~\ref{tab:head_budgets}); they cannot be pooled as one arm. No unattenuated trio met all initial matching constraints; the equal-norm adjacent-versus-distant locality criterion did not pass. The component result supports partial value transmission, without a magnitude-independent claim about spatial locality.

\paragraph{GELU hybrids.}
These local interventions separate two terms in the clean-noisy activation difference. Let $g$ denote GELU, and let $z_C,z_S$ be its clean and noisy preactivations. Define
\begin{align}
 L&=g'(z_C)(z_S-z_C),\\
 N&=g(z_S)-g(z_C)-L.
\end{align}
The first-order term $L$ is evaluated at the clean preactivation, and $N$ is the exact remainder. First-order reversal sets the activation to $g(z_S)-L=g(z_C)+N$; remainder reversal sets it to $g(z_S)-N=g(z_C)+L$. Full restoration inserts $g(z_C)$. The LayerNorm intervention applies clean statistics to noisy residual inputs.

\begin{figure}[t]
\centering\includegraphics[width=0.85\linewidth]{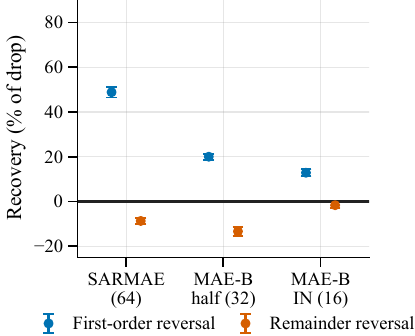}
\caption{\textbf{GELU hybrids on selected neurons.} Recovery divides by each model's corruption drop. Parentheses give the neuron counts; IN denotes ImageNet normalization. Intervals are 95\%. Table~\ref{tab:gelu_compact} also reports whole-layer interventions.}
\label{fig:gelu_supp}
\end{figure}

For the selected subsets, first-order reversal improves reconstruction and remainder reversal worsens it in SARMAE and both MAE-B preprocessing branches (Fig.~\ref{fig:gelu_supp}). Table~\ref{tab:gelu_compact} reports the selected-set and whole-layer results. Whole-layer first-order reversal recovers 92.6\% in SARMAE, compared with 5.9\% and 1.7\% in the two MAE-B branches. The large whole-layer effect is therefore specific to SARMAE among these comparisons.

The interventions have different write magnitudes and are followed by nonlinear computation. Their output effects do not form an additive decomposition of reconstruction error or establish that the model is globally linear. A planned analytic corrector required a whole-MLP nonlinear-rectification criterion. That criterion was not met, so the corrector was not run.

\begin{table*}[t]\centering\small
\caption{GELU hybrids: correlation-drop recovery (\%) with 95\% intervals. Selected sets contain 64, 32 and 16 neurons, respectively. Whole-layer targets are SARMAE MLP1 and MAE-B MLP3. LN uses clean LayerNorm statistics; full inserts clean post-GELU values. Each model/branch has its own corruption-drop denominator.}\label{tab:gelu_compact}
\begin{tabular}{@{}llrrr@{}}\toprule
Target & Intervention & SARMAE & MAE-B, half & MAE-B, ImageNet\\\midrule
Selected set & LN statistics & 9.7 [8.6,10.9] & -13.9 [-15.7,-12.2] & -15.7 [-17.6,-13.9]\\
 & Full restoration & 43.8 [41.4,46.2] & 18.0 [16.5,19.5] & 12.1 [10.7,13.5]\\
 & First-order reversal & 48.8 [46.4,51.2] & 20.0 [18.6,21.3] & 12.8 [11.3,14.3]\\
 & Remainder reversal & -8.7 [-10.1,-7.5] & -13.4 [-15.4,-11.3] & -1.7 [-2.8,-0.3]\\\midrule
Whole layer & LN statistics & 15.7 [13.5,17.9] & -49.8 [-55.5,-44.3] & -40.2 [-46.2,-34.8]\\
 & Full restoration & 82.1 [79.9,84.3] & 33.7 [30.7,36.8] & 24.5 [21.0,27.8]\\
 & First-order reversal & 92.6 [90.0,95.3] & 5.9 [2.2,9.4] & 1.7 [-3.1,5.8]\\
 & Remainder reversal & -45.9 [-53.6,-38.5] & -54.2 [-61.0,-47.7] & -70.3 [-78.5,-62.9]\\
\bottomrule\end{tabular}\end{table*}

\section{Second-checkpoint scope}\label{app:F}
MAE-B repeats discovery and confirmation with one checkpoint under the two unresolved normalizations. Both restoration profiles peak at MLP3; injection peaks at MLP1. At the selected sizes (32 for half, 16 for ImageNet), attribution outperforms sensitivity and three random sets (Table~\ref{tab:mae_sets}). Its equal-budget selected-minus-control contrasts remain positive, but the location result does not reproduce SARMAE's MLP1 maximum. Complete layer profiles, including negative effects, are in the accompanying data.
\begin{table*}[t]\centering\small
\caption{\textbf{Held-out recovery for selected and comparison sets.} Columns compare the two MAE-B normalizations; brackets give 95\% confidence intervals.}\label{tab:mae_sets}
\begin{tabular}{@{}>{\raggedright\arraybackslash}p{6.4cm}p{4.8cm}p{4.8cm}@{}}
\toprule Arm & Half (\%) [CI] & ImageNet (\%) [CI]\\\midrule

Attribution (selected size) & 18.0 [16.5, 19.5] & 12.1 [10.7, 13.5] \\
First random set & 0.7 [0.6, 0.9] & 0.1 [0.0, 0.2] \\
Second random set & 0.8 [0.6, 1.0] & 0.4 [0.3, 0.5] \\
Third random set & 0.9 [0.7, 1.0] & 0.0 [-0.1, 0.1] \\
Sensitivity & -1.1 [-2.1, 0.0] & 4.4 [3.6, 5.2] \\
\bottomrule\end{tabular}
\end{table*}

Confirmation clean/noisy correlations are 0.9820/0.7538 for half (drop 0.2282 [0.2040,0.2539]) and 0.9862/0.7643 for ImageNet (drop 0.2219 [0.1974,0.2478]). SARMAE's mechanism-confirmation values are 0.9468/0.6522 (drop 0.2946 [0.2712,0.3178]). Different training, preprocessing and output denormalization preclude attributing the location shift or raw differences to a single training ingredient.

\section{Real images and correction from noisy inputs}\label{app:G}
\paragraph{Manipulations and references.}
Sentinel-1 uses a no-data-aware $5\times5$ intensity average as its proxy, multiplied by single-look 1.5-pixel-grain speckle. Capella uses a $9\times9$ native-intensity average as its proxy; the manipulated input is $|h*z|^2$, with a unit-energy Gaussian complex-field filter of standard deviation 1.5 pixels, rescaled to the native window mean. The 174 Sentinel crops cover 16 sites/15 acquisitions; 70 valid Capella crops cover six regions/eight acquisitions. Capella crop 209 is excluded for constant hidden reference pixels.

Mean proxy/noisy/restored correlations are 0.6785/0.5193/0.5638 (Sentinel) and 0.7993/0.6229/0.6847 (Capella). Original site/acquisition-grouped recovery is 28.0 [24.9,31.3]\% and 35.0 [27.8,40.0]\%; later region-aware Capella uncertainty is [25.9,40.0]\%, and Sentinel site/acquisition sensitivities are [24.8,31.2]/[24.7,31.2]\%. The reference-width analysis has 243 crops, additionally excluding crop 210 for a constant box-13 reference. These estimates are not pooled with the 244 original crops. Sentinel's compact-size comparison and Capella injection remain unresolved; Capella gain lacks a supported denominator, and its box-5 size interval crosses the specified threshold. Decoder equal-budget comparisons remain positive on these reused crops.

\paragraph{Native and temporal-reference limits.}
The independent temporal mean excludes the test date and enforces registration/stability checks. Five of six sites provide 57 eligible crops. Reference-input/native-input correlations are 0.598 [0.559,0.652] and 0.578 [0.537,0.635]; their drop is 0.0199 [0.0137,0.0249]. This falls below the 0.05 minimum, so conditional native restoration was not run. Gain drop is 0.017 [$-0.007,0.034$]. Native-Capella damage also misses its criterion. A later experiment adds controlled speckle to these same temporal references: reference-donor restoration recovers 27.8 [21.3,35.3]\%, while the non-oracle alternatives are in Table~\ref{tab:correction_compact}.

\paragraph{Correction methods and outcomes.}
Clean-donor restoration identifies targets with access to clean activations. The following corrections test these targets when only noisy observations are available at test time.
Ridge correctors predict clean post-GELU states from noisy activations and patch statistics. They fit 216 synthetic scenes; scene-grouped cross-validation selects regularization by activation MSE. Distributed correctors are fitted sequentially with earlier corrections applied. This correction experiment uses \emph{all 38 characterized selective neurons}, distinct from the 36-unit masked restoration/hold. Fits remain unchanged on confirmation and real images; mixed clean/noisy training was not used. Filtered donors take activations from a box-filtered observation; filters operate on linear intensity~\cite{lee1980filter}.
\begin{table*}[t]\centering\small
\caption{Non-oracle reconstruction correction, with 95\% intervals. The synthetic columns use the mechanism-confirmation cohort; cost is clean-input correlation loss. The last column adds controlled speckle to independent temporal references (57 reused crops). Each recovery divides by its own cohort's corruption drop.}\label{tab:correction_compact}
\begin{tabular}{@{}lrrr@{}}\toprule
Intervention & Synthetic recovery (\%) & Synthetic clean cost & Temporal-reference recovery (\%)\\\midrule
64 restoration-neuron correction & 20.6 [18.4,22.7] & 0.0165 [0.0119,0.0217] & 16.7 [10.5,23.6]\\
38 selective-neuron correction & -3.7 [-6.0,-1.2] & 0.0006 [0.0003,0.0009] & -4.4 [-6.7,-2.4]\\
Whole-MLP1 correction & 44.4 [39.8,48.3] & 0.0923 [0.0787,0.1070] & 30.8 [21.8,40.7]\\
Box filter, $7\times7$ & 47.6 [45.1,50.1] & 0.0094 [0.0070,0.0119] & 49.2 [42.8,54.9]\\
Lee filter, $7\times7$ & 45.9 [43.4,48.5] & 0.0094 [0.0070,0.0119] & 50.0 [42.9,56.3]\\
Whole-MLP1 filtered donor & 46.9 [44.1,49.7] & 0.0053 [0.0034,0.0074] & 41.0 [35.4,45.2]\\
\bottomrule\end{tabular}\end{table*}

On mechanism confirmation, restoration-neuron minus selective-neuron correction is 24.3 [19.7,28.6] recovery points; whole-MLP correction minus box filtering is unresolved, $-3.2$ [$-9.7,2.6$]. On the temporal-reference follow-up, whole-MLP filtered donor minus box filtering is $-8.3$ [$-13.8,-1.2$] points (97.5\% interval). In the earlier additional confirmation cohort, 64-neuron insertion, whole-MLP insertion and full filtering recover 30.1\%, 46.0\% and 50.5\%, respectively; targeted-minus-full is negative.

\paragraph{Scope.}
Smoothed references favor full-input filtering and are not clean physical reflectivity. Small native damage and clean-input correction costs constrain operational claims. These reused-data tests do not verify disjointness from training geography. Together with the simple synthetic scenes and two ViT-B checkpoints, they support a partial, conditional computation account, not an exhaustive or universally transferable circuit.

\section{Recognition methods and boundaries}\label{app:H}
\paragraph{Data, preprocessing and readout.}
Recognition uses BigEarthNet-S1 v2 VV and all 19 land-cover labels~\cite{sumbul2021bigearthnetmm,clasen2025reben}: 4,800/1,600/1,600 training/validation/test patches from 32/10/10 observed MGRS tiles. Tile assignments use metadata-only label/country balancing; repeated dates remain in one split, and cross-split overlapping footprints are excluded before feature extraction. The grouping is by tile, not the stricter acquisition-linked grouping.

Stored dB values are converted to linear intensity, resized bilinearly from $120\times120$ to $224\times224$, then multiplied by a fixed single-look 1.5-pixel-grain speckle realization. Training pixels alone set gain 1.32017, mapping their median to $-10$ dB within the fixed display window. Images are repeated over three channels with half normalization. All 196 patches are visible. The readout averages final-normalized patch tokens without CLS; neuron interventions include all 197 tokens.

Nineteen L2 logistic regressions fit native training features without additional fitted scaling. Native-validation macro-AP selects shared $C$ from 13 log-spaced values $10^{-4}$-$10^2$ (ties choose smaller); $C=1$ is selected. L-BFGS tolerance is $10^{-6}$, with up to 20,000 iterations for unfinished fits. Per-class native-validation F1 thresholds use 0.05-0.95 in 0.01 steps, ties choosing higher; these affect secondary micro-F1, not AP.

\paragraph{Uncertainty.}
Ten thousand paired bootstrap draws resample the ten test tiles, including all their patches; estimates pool patches. A label absent in a draw contributes AP zero rather than being omitted (the most affected label is absent in 3,516 draws). Intervals condition on the fitted model/probe, observed tiles and fixed noise realization. Recovery is $(\mathrm{AP}_G-\mathrm{AP}_S)/(\mathrm{AP}_N-\mathrm{AP}_S)$; injection damage is $(\mathrm{AP}_N-\mathrm{AP}_G)/(\mathrm{AP}_N-\mathrm{AP}_S)$. Ratios are recomputed per draw. Raw AP percentage points and percentages of the AP drop are different units.

\paragraph{Non-oracle interventions.}
Table~\ref{tab:recognition_baselines_compact} retains every native/speckled correction and filtering condition. The synthetic-trained whole-MLP corrector changes speckled AP by 1.51 [$-0.31,2.99$] points; filtered-donor insertion, a secondary comparison that uses the filtered observation only, gains 3.63 points. Full-input filters perform better, and native-input costs remain substantial.

\begin{table}[t]
\centering
\caption{Frozen-probe recognition: all native and added-speckle conditions.
Macro-AP (\%) with paired 95\% tile-bootstrap intervals.
The same synthetic-trained corrector and fixed filters are used without refitting.}
\label{tab:recognition_baselines_compact}
\scriptsize
\setlength{\tabcolsep}{2.5pt}
\begin{tabular}{@{}p{0.5\columnwidth}rr@{}}
\toprule
Intervention & Native input & Added speckle \\
\midrule
None
& 47.75 [42.54,49.72]
& 28.22 [26.05,32.46] \\

Whole-MLP1 learned correction
& 32.78 [29.73,36.69]
& 29.73 [27.41,33.78] \\

Box filter
& 35.24 [32.73,38.34]
& 37.09 [33.57,40.71] \\

Lee filter
& 35.32 [32.80,38.47]
& 37.34 [34.06,40.62] \\

Whole-MLP1 filtered donor
& 40.16 [37.03,43.29]
& 31.85 [29.87,35.08] \\
\bottomrule
\end{tabular}
\end{table}

\begin{table*}[!b]
\centering
\footnotesize
\setlength{\tabcolsep}{4pt}
\caption{Native-donor recognition interventions: percentages of the
native-minus-speckle AP drop, with pointwise 95\% tile-bootstrap intervals.
Restoration measures improvement over speckled input; injection measures
deterioration from native input. The 38 selective neurons are the unmasked
set. Dashes denote unmeasured combinations.}
\label{tab:recognition_interventions_compact}

\begin{tabular}{@{}lrr@{}}
\toprule
Target & Restoration recovery (\%) & Injection damage (\%)\\
\midrule
Whole MLP1 & 21.9 [14.7,30.8] & 100.4 [92.7,115.5]\\
64 restoration neurons & -1.0 [-8.6,5.1] & 14.6 [5.3,20.0]\\
64 sensitivity-ranked neurons & -1.1 [-8.3,3.4] & 39.4 [27.4,46.9]\\
64 activation-ranked neurons & 1.1 [-3.2,4.6] & -\\
38 selective neurons & 4.8 [-4.4,9.7] & 31.7 [19.6,40.4]\\
Random set 1, 64 neurons & -0.0 [-0.3,0.6] & 0.8 [-1.2,3.1]\\
Random set 2, 64 neurons & -0.2 [-0.8,0.7] & -\\
Random set 3, 64 neurons & 0.6 [-2.1,2.8] & -\\
Layer-count control 1 & 0.6 [-0.1,1.9] & -\\
Layer-count control 2 & 1.3 [0.4,2.7] & -\\
Layer-count control 3 & 0.0 [-0.8,0.7] & -\\
Wrong donor, 64 restoration neurons & -6.9 [-20.0,-1.6] & -\\
Wrong donor, whole MLP1 & -22.3 [-44.2,-13.7] & -\\
\midrule
\multicolumn{3}{l}{Size sweep: restoration recovery (\%), same pointwise 95\% intervals}\\
\midrule
Size & Attribution / sensitivity & Random permutation\\
\midrule
8 & -0.2 [-1.6,1.4] / 0.2 [-5.1,3.8] & 0.2 [-0.4,0.9]\\
16 & -0.7 [-3.8,1.8] / -0.1 [-5.8,4.1] & 0.2 [-0.5,0.9]\\
32 & -0.3 [-6.3,3.9] / -0.8 [-7.3,3.9] & 0.1 [-1.0,1.6]\\
128 & -1.3 [-10.7,8.0] / -2.0 [-11.8,3.7] & 0.7 [-1.4,3.0]\\
256 & 2.4 [-7.4,12.5] / -4.7 [-16.8,3.6] & 2.2 [0.2,4.4]\\
\bottomrule
\end{tabular}
\end{table*}
\paragraph{Native-donor restoration and injection.}
These tests ask whether the reconstruction targets also recover recognition performance. Donors are the unaugmented native SAR observations. The checkpoint, probe, examples, noise, readout and order remain fixed. Native-donor restoration inserts native-image activations into a speckled pass; injection reverses this operation. Whole-layer scans cover all 12 MLPs. Sparse sets include the 64 restoration neurons, same-layer sensitivity/activation rankings, three random sets, and all 38 unmasked speckle-selective neurons with three layer-count controls. Wrong donors use another test image. Table~\ref{tab:recognition_interventions_compact} retains these comparisons and the full sparse size sweep. Unpatched score reproduction and self-patches are checked; pointwise scan intervals are not multiplicity-adjusted evidence for a uniquely best layer.

MLP1 has the largest restoration point estimate: 4.28 [2.47,5.65] AP points, or 21.9 [14.7,30.8]\% of the drop. MLP7 and MLP12 recover 17.4 [10.7,24.8]\% and 12.2 [8.0,16.5]\%; MLP10 harms performance, $-7.3$ [$-19.0,-0.9$]\%. The 64 restoration neurons change AP by $-0.19$ [$-1.37,0.88$] points. Their recovery-point contrasts against mean random, sensitivity and selective sets are $-1.09$ [$-8.09,4.56$], 0.14 [$-4.28,5.69$] and $-5.80$ [$-14.05,4.84$]. Selective minus layer-count controls is 4.20 [$-5.21,8.94$]. The random 256-neuron sweep has a small positive result; no general impossibility of sparse repair follows.

\paragraph{Training-selected neurons and ablations.}
This exploratory follow-up reuses the already-inspected test cohort. Training examples alone rank first-MLP units by
\begin{equation}
s_j=-\frac1n\sum_i\sum_t\frac{\partial\ell_i(h_S)}{\partial h_{S,itj}}(h_{N,itj}-h_{S,itj}),
\end{equation}
where $\ell$ averages binary cross-entropy across 19 labels and the token sum includes CLS. Signed scores descend, ties use coordinate index; the top 64 are fixed without a positivity requirement. This BCE surrogate does not guarantee AP improvement. The 4,800 native training images also provide per-coordinate means over images and tokens. The selected set shares five units with the reconstruction set.

Zero/mean ablations cross native/speckled input with the 38 selective neurons and the three layer-count controls. Table~\ref{tab:recognition_followup_compact} reports every input-matched effect and all seven adjusted primary comparisons. Training-selected restoration gives $-0.27$ [$-7.51,3.96$]\% recovery (95\%). None of the seven comparisons excludes zero at either its pointwise 95\% or adjusted 99.2857\% level; unresolved native-input changes are not equivalence tests.
\begin{table*}[t]\centering\small
\caption{Training-selected restoration and selective-neuron ablation. Upper panel: AP changes in percentage points from the input-matched baseline, with pointwise 95\% intervals; means use native training images only. Lower panel: all seven primary comparisons with 99.2857\% Bonferroni intervals. AP p.p. and recovery p.p. are distinct units.}\label{tab:recognition_followup_compact}
\begin{tabular}{@{}lrr@{}}\toprule
Replacement / target & Native-input AP change & Speckled-input AP change\\\midrule
Zero / 38 selective neurons & -0.03 [-0.59,0.31] & 1.18 [-0.51,2.04]\\
Zero / control 1 & 0.06 [-0.08,0.25] & 0.10 [-0.03,0.22]\\
Zero / control 2 & -0.46 [-0.62,-0.06] & 0.00 [-0.20,0.19]\\
Zero / control 3 & -0.03 [-0.18,0.07] & 0.08 [-0.02,0.17]\\
Mean / 38 selective neurons & -0.19 [-0.32,0.11] & 1.13 [-0.53,1.98]\\
Mean / control 1 & -0.00 [-0.04,0.09] & 0.05 [-0.07,0.18]\\
Mean / control 2 & -0.13 [-0.31,-0.02] & 0.12 [-0.04,0.29]\\
Mean / control 3 & 0.02 [-0.13,0.13] & 0.07 [-0.04,0.18]\\
Training-selected 64, native donor & - & -0.05 [-1.26,0.70]\\\midrule
Primary comparison & Units & Estimate [adjusted interval]\\\midrule
Training-selected set: recovery & recovery p.p. & -0.27 [-11.06,5.81]\\
Training-selected minus random set & recovery p.p. & -0.39 [-11.38,6.53]\\
Training-selected minus restoration set & recovery p.p. & 0.70 [-11.66,15.34]\\
Zero replacement: selective-set gain & AP p.p. & 1.18 [-1.17,2.32]\\
Zero replacement: selective minus controls & AP p.p. & 1.12 [-1.17,2.24]\\
Mean replacement: selective-set gain & AP p.p. & 1.13 [-1.24,2.28]\\
Mean replacement: selective minus controls & AP p.p. & 1.05 [-1.27,2.18]\\
\bottomrule\end{tabular}\end{table*}

\paragraph{Training comparison.}
Rank-8 LoRA updates both linear maps of every encoder MLP and trains a classifier, leaving base weights fixed~\cite{hu2021lora}; every arm has 751,891 trainable parameters. Loss combines native/speckled BCE and an optional consistency penalty normalized by its initial training-pair mean. Neuron penalties include CLS; feature consistency uses the pooled output. Augmentation-only validation selects learning rate $3\times10^{-4}$ and five epochs for all arms. AdamW uses weight decay 0.01, $\beta=(0.9,0.999)$, $\epsilon=10^{-8}$ and batches of 32 pairs. Speckled-validation AP selects penalty weight from $\{0.1,0.3,1,3,10\}$, separately for each of three random sets. Three paired seeds share initialization and minibatch order; AP is computed per fitted model before averaging over sets and seeds.
\begin{table}[t]
\centering
\scriptsize
\setlength{\tabcolsep}{2.5pt}
\caption{Low-rank training comparisons. AP is computed per model, then averaged over sets within seed and over three seeds. Intervals are 95\% tile-bootstrap intervals, conditional on the fitted models.}
\label{tab:lowrank_training}

\begin{tabular}{@{}p{0.38\columnwidth}rr@{}}
\toprule
Training condition & Native macro-AP (\%) & Speckled macro-AP (\%)\\
\midrule
Augmentation only
& 48.60 [43.37,50.61]
& 44.71 [40.69,47.47]\\

Pooled-feature consistency
& 47.57 [42.63,50.79]
& 45.75 [41.34,48.58]\\

Restoration-neuron consistency
& 48.37 [43.18,50.82]
& 44.83 [40.69,47.57]\\

Speckle-selective consistency
& 48.96 [43.86,50.86]
& 45.06 [41.13,47.43]\\

Random-neuron consistency
& 48.20 [42.95,50.81]
& 44.68 [40.68,47.33]\\

Filtered-input training
& 46.35 [41.43,49.68]
& 45.24 [40.49,47.92]\\
\bottomrule
\end{tabular}
\end{table}

The two primary 97.5\% AP-point contrasts are unresolved: restoration-neuron minus selective-neuron consistency is $-0.23$ [$-0.75,0.50$], and minus feature consistency is $-0.92$ [$-1.83,0.08$]. Secondary 95\% contrasts against augmentation and filtered training are 0.12 [$-0.26,0.46$] and $-0.41$ [$-0.97,0.71$]. The separate equal-tile estimand against random consistency is $-0.29$ [$-0.65,0.05$]; each random set chooses weight 10. Filtered training costs 2.25 [0.65,2.91] native AP points relative to augmentation. The 44.71 versus 28.22 speckled-AP difference between augmentation and the frozen probe compares differently fitted models. Images, visible-patch count, pooling/readout and objective all change from reconstruction; these experiments therefore bound transfer across that setting change, rather than identifying the objective alone as its cause.

\section{Reproduction}
All experiments use the publicly released SARMAE checkpoint and the ImageNet MAE-B visualization checkpoint.
The synthetic scenes are generated from the procedures in Appendix~B, and the real-image and recognition experiments use the public Sentinel-1, Capella, and BigEarthNet-S1 data described in Appendices~G and~H.


\end{document}